%% file: article.tex
\documentclass{article}

\usepackage{arxiv}

\usepackage[utf8]{inputenc} 
\usepackage[T1]{fontenc}    
\usepackage{url}            
\usepackage{booktabs}       
\usepackage{amsmath,amsfonts,amssymb}
\usepackage{nicefrac}       
\usepackage{microtype}      
\usepackage{changes}
\usepackage{amsmath}
\usepackage{graphicx}
\usepackage{hyperref}       
\usepackage{cleveref}
\usepackage{algorithm}
\usepackage{algpseudocode}
\usepackage{natbib}
\usepackage[separate-uncertainty=true]{siunitx}
\usepackage{wrapfig}
\usepackage[]{charter}
\setcitestyle{numbers}

\newcommand{\taum}{\tau_\textrm{mem}}
\newcommand{\taus}{\tau_\textrm{syn}}

\newcommand{\taupost}[1]{\frac{\mathrm{d}\tpost{#1}}{\mathrm{d}w_{ji}}}
\newcommand{\tpost}[1]{t^{\textrm{post}}_{#1}}

\newcommand{\npost}{N_{\textrm{post}}}

\newcommand{\ddt}{\frac{\mathrm{d}}{\mathrm{d}t}}

\newcommand{\dt}{\mathrm{d}t}

\begin{document}
\title{Event-Based Backpropagation can compute Exact Gradients for Spiking Neural Networks}
\author{Timo C. Wunderlich\thanks{The authors have contributed equally.} \\
    Kirchhoff-Institute for Physics\\
    Heidelberg University\\
    69120 Heidelberg, Germany \\
    \emph{Current Address:} \\
    Berlin Institute of Health\\
    Charité–Universitätsmedizin\\
    10117 Berlin, Germany\\
    \texttt{timo.wunderlich@charite.de}
    \And
    Christian Pehle\footnotemark[1]\\
    Kirchhoff-Institute for Physics\\
    Heidelberg University\\
    69120 Heidelberg, Germany \\
    \texttt{christian.pehle@kip.uni-heidelberg.de}} \maketitle \begin{abstract}
    Spiking neural networks combine analog computation with event-based communication using discrete spikes.
    While the impressive advances of deep learning are enabled by training non-spiking artificial neural networks using the backpropagation algorithm, applying this algorithm to spiking networks was previously hindered by the existence of discrete spike events and discontinuities.
    For the first time, this work derives the backpropagation algorithm for a continuous-time spiking neural network and a general loss function by applying the adjoint method together with the proper partial derivative jumps, allowing for backpropagation through discrete spike events without approximations.
    This algorithm, EventProp, backpropagates errors at spike times in order to compute the exact gradient in an event-based, temporally and spatially sparse fashion.
    We use gradients computed via EventProp to train networks on the Yin-Yang and MNIST datasets using either a spike time or voltage based loss function and report competitive performance.
    Our work supports the rigorous study of gradient-based learning algorithms in spiking neural networks and provides insights toward their implementation in novel brain-inspired hardware.
\end{abstract}

\pagebreak
\begin{figure}
    \centering
    \includegraphics[width=.6\textwidth]{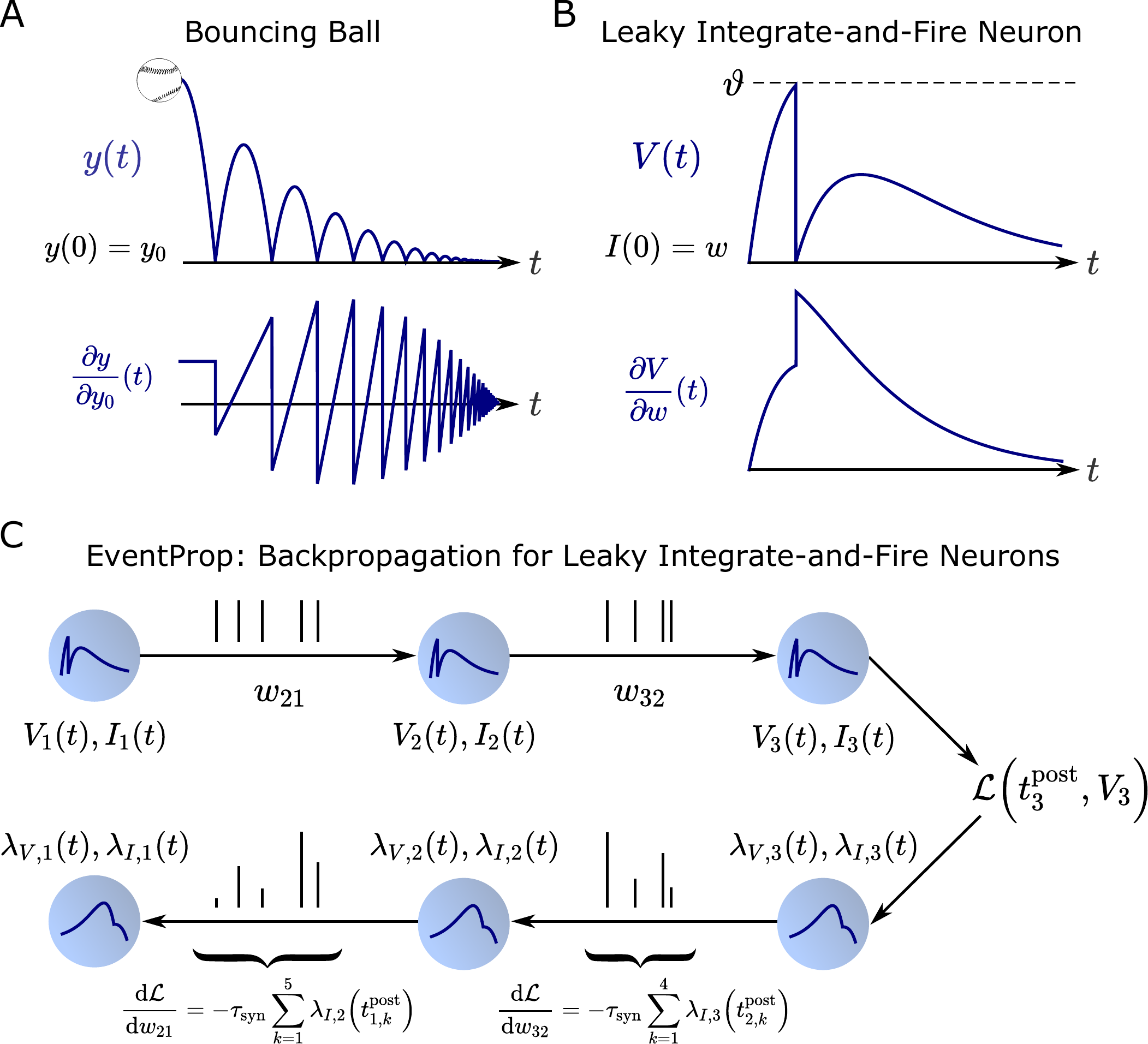}
    \caption{We derive the precise analogue to backpropagation for spiking neural networks by applying the adjoint method together with the jump conditions for partial derivatives at state discontinuities, yielding exact gradients with respect to loss functions based on membrane potentials or spike times.
    \textbf{A, B}: Dynamical systems with parameter-dependent discontinuous state transitions typically have discontinuous partial derivatives of state variables with respect to system parameters \cite{Barton2002}, as is the case for the two examples shown here.
    Both examples model dynamics occurring on short timescales, namely inelastic reflection and the neuronal spike mechanism, using an instantaneous state transition.
    We denote quantities evaluated before and after a given transition by $-$ and $+$.
    In A, a bouncing ball starts at height $y_0>0$ and is described by $\ddot y=-g$ with gravitational acceleration $g$.
    It is inelastically reflected as $\dot y^+=-0.8\dot y^-$ as soon as $y^-=0$ holds, causing the partial derivative with respect to $y_0$ to jump as $\frac{\partial y^+}{\partial y_0}=-0.8\frac{\partial y^-}{\partial y_0}$ (see \cref{sec:pdh}).
    In B, a leaky integrate-and-fire neuron described by the system given in \cref{fig:lif} with initial conditions $I(0)=w$, $V(0)=0$ resets its membrane potential as $V^+=0$ when $V^-=\vartheta$ holds, causing the partial derivative to jump as $\frac{\partial V^+}{\partial w}=\left(\frac{\vartheta}{\taum \dot V^-}+1\right)\frac{\partial V^-}{\partial w}$ (see \cref{sec:control-derivation}).
    \textbf{C}: Applying the adjoint method with partial derivative jumps to a network of leaky integrate-and-fire neurons (\cref{fig:lif}) yields the adjoint system (\cref{fig:adjoint-system}) that backpropagates errors in time.
    EventProp is an algorithm (\cref{alg:eventprop}) returning the gradient of a loss function with respect to synaptic weights by computing this adjoint system.
    The forward pass computes the state variables $V(t)$, $I(t)$ and stores spike times $\tpost{}$ and each firing neuron's synaptic current.
    EventProp then performs the backward pass by computing the adjoint system backwards in time using event-based error backpropagation and gradient accumulation: each time a spike was transferred across a given synaptic weight in the forward pass, EventProp backpropagates the error signal represented by the adjoint variables $\lambda_V(\tpost{})$, $\lambda_I(\tpost{})$ of the post-synaptic (target) neuron and updates the corresponding component of the gradient by accumulating $\lambda_I(\tpost{})$, finally yielding sums as given in the figure.
     }\label{fig:fig1}
\end{figure}
\section{Introduction}
How can we train spiking neural networks to achieve brain-like performance in machine learning tasks?
The resounding success and pervasive use of the backpropagation algorithm in deep learning suggests an analogous approach.
This algorithm computes the gradient of the neural network parameters with respect to a loss function that measures the network's performance in a given task.
The parameters of the network are iteratively updated using the locally optimal direction given by the gradient.

Spiking neural networks have been referred to as the third generation of neural networks  \citep{Maass1997}, superseding artificial neural networks as commonly used in deep learning and hold the promise for efficient and robust processing of event-based spatio-temporal data as found in biological systems.
However, spiking models are not widely used in machine learning applications.
At the same time, the development of spiking neuromorphic hardware receives increasing attention \citep{Kaushik2019} and learning in spiking neural networks is an active research subject, with  a wide variety of proposed algorithms.
A notorious issue in spiking neurons is the hard spiking threshold that does not permit a straight-forward application of differential calculus to compute gradients.
Although exact gradients have been derived for special cases, this issue is commonly side-stepped by using smoothed or stochastic neuron models or by replacing the hard threshold function using a surrogate function, leading to the computation of surrogate gradients \citep{neftci:2019}.

In contrast, this work provides an algorithm, EventProp, to compute the exact gradient for an arbitrary loss function defined using the state variables (spike times and membrane potentials) of a general recurrent spiking neural network composed of leaky integrate-and-fire neurons with hard thresholds.
Since feed-forward architectures correspond to recurrent neural networks with block weight matrices and convolutions can be represented as sparse linear transformations, deep feed-forward networks and convolutional networks are included as special cases.

\subsection*{Partial Derivatives in Discontinuous Dynamical Systems}
The leaky integrate-and-fire neuron model describes a hybrid dynamical system that combines continuous dynamics between spikes with discontinuous state variable transitions at spike times.
The computation of partial derivatives for hybrid dynamical systems is an established topic in optimal control theory \citep{Barton2002,rozenwasser2019sensitivity}.
In hybrid systems, the time-dependent partial derivative $\frac{\partial x}{\partial p}(t)$ of a state variable $x$ with respect to a parameter $p$ generally experiences jumps at the points of discontinuity (see \cref{fig:fig1} A, B).
The relation between the partial derivatives before and after a given discontinuity was first studied in the 1960s \citep{DEBACKER1964168,rozenvasser:1967}.
A more general theoretical framework was developed thirty years later \cite{GALAN199917}, providing existence and uniqueness theorems for the partial derivative trajectories $\frac{\partial x}{\partial p}(t)$ of hybrid systems.

Discontinuous state transitions in hybrid systems occur when a transition condition is fulfilled (e.g., a bouncing ball hits the floor or a neuron reaches its spiking threshold).
The existence of well-defined partial derivative jumps at the state transition times depends on the local applicability of the implicit function theorem to the transition condition, requiring that the event time depends on the parameters in a differentiable fashion.
In the case considered here, a spiking neural network composed of leaky integrate-and-fire neurons that is parameterized by synaptic weights, this is fulfilled up to the null set in weight space that contains the locally defined hypersurfaces where spikes are added or removed.
At these critical points, the derivative of the time of the (dis-)appearing spike with respect to a given active synaptic weight diverges.
This implies that both the spike times and an integral of a smooth loss function over the membrane potential are differentiable almost everywhere, up to the null set of critical points in weight space.

\subsection*{Backpropagation of Errors in Discontinuous Dynamical Systems}
Having established the jumps of partial derivatives in the leaky integrate-and-fire neuron model, the relevant question is how to compute the gradient of a loss function for spiking neural networks, preferably with the computational efficiency afforded by the backpropagation algorithm and retaining any potential advantages of event-based communication.
Backpropagation in discrete-time artificial neural networks can be derived as a special case of the adjoint method \cite{LeCun1988ABack-propagation}, with the adjoint variables (Lagrange multipliers) $\lambda_t$ at each time step $t$ corresponding to the intermediate variables computed in the backpropagation algorithm.
Applying the adjoint method to continuous-time dynamical systems yields time-dependent adjoint variables $\lambda(t)$ (see \cref{sec:adjoint}) and their computation in reverse time is analogous to the backpropagation of errors in discrete-time artificial neural networks.
The adjoint method can be applied to hybrid systems by using the proper partial derivative jumps that generally cause jumps in the adjoint variables \cite{Serban_2019}.

\subsection*{EventProp: Event-Based Backpropagation of Errors}
We combine the partial derivative jumps of the leaky integrate-and-fire neuron with the adjoint method in order to derive the EventProp algorithm (\cref{alg:eventprop}) that is the analogue to backpropagation for spiking neural networks (\cref{fig:fig1} C).
Since EventProp backpropagates errors at spike times, the algorithm computes gradients using an event-based communication scheme and is amenable to neuromorphic implementation.
By requiring the storage of state variables only at spike times, it provides favorable memory requirements compared to approaches that require the full forward state trajectory to be retained for the backward pass.
For example, surrogate gradient approaches operating on a discrete time grid typically require storing state variables at every time step for the backward pass \citep[however, an effort to compute surrogate gradients in a more sparse fashion has been made ][]{pereznieves2021sparse}.
More generally, the fact that backpropagation in discrete-time artificial neural networks requires storing activations at every time step causes a memory bottleneck and is a major concern in training very deep architectures \cite[e.g., ][]{PleissCHLMW17,NEURIPS2019_ffe10334,bs-2003-08732}.

EventProp does not prescribe a specific numerical scheme to compute state variables and spike times but since the backward pass corresponds to the computation of a spiking network with pre-determined spike times, the computational complexity of the backward pass generally corresponds to that of the forward pass.
While surrogate gradient approaches on a discrete time grid typically require the calculation of dense matrix-vector products at every time step in the backward pass (all neurons backpropagate error signals at every time step), EventProp only requires computing vector-vector products at spike events (only the firing neuron receives backpropagated errors at a given spike time).
In this way, EventProp leverages the sparseness of spike-based communication for both the forward and backward pass.

We demonstrate the training of spiking neural networks with a single hidden layer using EventProp and the Yin-Yang and MNIST datasets, resulting in competitive classification performance.

\subsection*{Previous Work}
We refer the reader to the following review articles for a comprehensive survey of gradient-based approaches to learning in spiking neural networks: \cite{Pfeiffer2018} and \cite{TAVANAEI201947} discuss learning in deep spiking networks, \cite{Kaushik2019} discuss learning along with the history and future of neuromorphic computing and \cite{neftci:2019} focus on the surrogate gradient approach.
Surrogate gradients use smooth activation functions for the purposes of backpropagation and have been used to train spiking networks in a variety of settings \citep[e.g., ][]{esser:2016,bellec:2018, zenke:2018, shrestha:2018slayer}.
This approach is typically derived by considering the Euler discretization of a spiking neural network where the Heaviside step function is used to couple neurons across discrete time steps.
The non-differentiable Heaviside step function is then replaced by a smooth function in the backward pass.

Apart from surrogate gradients, several publications provide exact gradients for first-spike-time based loss functions and leaky integrate-and-fire neurons: \cite{bohte:2000spikeprop} provides the gradient for at most one spike per layer and this result was subsequently generalized to an arbitrary number of spikes as well as recurrent connectivity \citep{Booij2005,Yan2013}.
While these publications provide recursive relations for the gradient that can be implicitly computed using backpropagation, we explicitly provide the dynamical system that implements backpropagation through time and show that it represents an adjoint spiking network which transmits errors at spike times, allowing for an event-based computation of the gradient.
In addition, we also consider voltage-dependent loss functions and our methodology can be applied to neuron models without analytic expressions for the post-synaptic potential kernels.

The applicability of methods from optimal control theory (i.e., partial derivative jumps and the adjoint method) to compute exact gradients in hard-threshold spiking neural networks was recognized in a series of publications \cite{ueyama2010,kuroe2006,mori2000}.
In contrast to this work, these articles consider a neuron model with a two-sided threshold (including negative threshold crossings), rely on the existence of analytic expressions for the post-synaptic potential kernels, provide specialized algorithms tailored to specific loss functions and consider minimalistic regression tasks.

The chronotron \citep{Florian2012} uses a gradient-based learning rule based on the Victor-Purpura metric which enables a single leaky integrate-and-fire neuron to learn a target spike train.
Our work, as well as the works mentioned above which derive exact gradients, applies the implicit function theorem to differentiate spike times with respect to synaptic weights.
A different approach is to consider ratios of the neuronal time constants where analytic expressions for first spike times can be given and to derive the corresponding gradients, as done in \cite{gltz2019fast,Comsa2020,Mostafa_2017,Kheradpisheh2020}.
Our work encompasses the contained methods to compute the gradient as special cases.

The seminal Tempotron model uses gradient descent to adjust the sub-threshold voltage maximum in a single neuron \citep{Gutig2006} and has recently been generalized to the spike threshold surface formalism \citep{Gutigaab4113} that uses the exact gradient of the critical thresholds $\vartheta^*_k$ at which a leaky integrate-and-fire neuron transitions from emitting $k$ to $k-1$ spikes; computing this gradient is not considered in this work.
The adjoint method was recently used to optimize neural ordinary differential equations \citep{chen:2018neural} and neural jump stochastic differential equations \citep{jia:2019neural} as well as to derive the gradient for a smoothed spiking neuron model without reset \citep{dongsung:2018}.

\clearpage
\section{Results}
We first define the used spiking neuron model and then proceed to state our main results.
\subsection{Leaky Integrate-and-Fire Neural Network Model}
\label{sec:lif}
We define a network of $N$ leaky integrate-and-fire neurons with arbitrary (up to self-connections) recurrent connectivity (\cref{fig:lif}).
We set the leak potential to zero and choose parameter-independent initial conditions.
Note that the Spike-Response Model (SRM) \citep{gerstner2002spiking} with double-exponential or $\alpha$-shaped PSPs is generally an integral expression of the model given in \cref{fig:lif} with corresponding time constants.
\begin{table}[h!]
    \centering
    \begin{tabular}{lll}
        \textbf{Free Dynamics} & \textbf{Transition Condition} & \textbf{Jumps at Transition} \\ \hline
        \rule{0pt}{2.8\normalbaselineskip}
        $\begin{aligned}
        \taum \ddt V &= -V + I \\
        \taus \ddt I &= -I
        \end{aligned}$&
        $\begin{aligned}
        (V)_n - \vartheta&=0\\
        (\dot V)_n &\neq 0\\
        \textrm{for any }&n
        \end{aligned}$
        &
        $\begin{aligned}
        (V^+)_n &= 0 \\
        I^+ &= I^- + We_n
        \end{aligned}$
        \\
    \end{tabular}
    \vspace{.3cm}
    \caption{The leaky integrate-and-fire spiking neural network model.
        Inbetween spikes, the vectors of membrane potentials $V$ and synaptic currents $I$ evolve according to the free dynamics.
        When some neuron $n\in [1..N]$ crosses the threshold $\vartheta$, the transition condition is fulfilled, causing a spike.
        This leads to a reset of the membrane potential as well as post-synaptic current jumps.
        $W\in \mathbb{R}^{N\times N}$  is the weight matrix with zero diagonal and $e_n\in \mathbb{R}^N$ is the unit vector with a $1$ at index $n$ and $0$ at all other indices.
        We use $-$ and $+$ to denote quantities before and after a given spike.}\label{fig:lif}
\end{table}

\subsection{Gradient via Backpropagation}
Consider smooth loss functions $l_V(V, t)$, $l_\mathrm{p}(\tpost{})$ that depend on the membrane potentials $V$, time $t$ and the set of post-synaptic spike times $\tpost{}$.
The total loss is given by
\begin{align}\label{eq:loss-basic}
    \mathcal{L} = l_\mathrm{p}(\tpost{})+\int_0^T l_V(V(t), t)\dt.
\end{align}
Our main result is that the derivative of the total loss with respect to a specific weight $w_{ji}=(W)_{ji}$ that connects pre-synaptic neuron $i$ (the firing neuron) to post-synaptic neuron $j$  (the receiving neuron) is given by a sum over the spikes caused by $i$,
\begin{align}\label{eq:loss-gradient}
    \frac{\mathrm{d}\mathcal{L}}{\mathrm{d}w_{ji}} = - \taus \sum_{\textrm{spikes from }i} (\lambda_I)_j,
\end{align}
where $\lambda_I$ is the adjoint variable (Lagrange multiplier) corresponding to the synaptic current $I$.
\Cref{eq:loss-gradient} therefore samples the post-synaptic neuron's adjoint variable $(\lambda_I)_j$ at the spike times caused by neuron $i$.

After the neuron dynamics given by \cref{fig:lif} have been computed from $t=0$ to $t=T$, the adjoint state variable $\lambda_I$ is computed in reverse time (i.e., from $t=T$ to $t=0$)  as the solution of the system of adjoint equations defined in \cref{fig:adjoint-system}.
The dynamical system defined by \cref{fig:adjoint-system} is the adjoint spiking network to the leaky integrate-and-fire network (\cref{fig:lif}) which backpropagates error signals at the spike times $\tpost{}$.

\begin{table}[h!]
    \centering
    \begin{tabular}{p{3.5cm}p{2cm}l}
        \textbf{Free Dynamics} & \textbf{Transition Condition} & \textbf{Jump at Transition} \\ \hline
        \rule{0pt}{3.3\normalbaselineskip}$\begin{aligned}
        \taum  \lambda_V' &= -\lambda_V - \frac{\partial l_V}{\partial V} \\
        \taus  \lambda_I' &= -\lambda_I + \lambda_V
        \end{aligned}$ &
        $\begin{aligned}
        t-\tpost{k} &= 0 \\
        \textrm{for any }&k
        \end{aligned}$ &
        $\begin{aligned}
        (\lambda_V^-)_{n(k)} &= (\lambda_V^+)_{n(k)} +  \frac{1}{\taum (\dot V^-)_{n(k)}} \bigg[\vartheta(\lambda_V^+)_{n(k)} \\
        &\quad+ \left(W^\top(\lambda_V^+ - \lambda_I)\right)_{n(k)}+ \frac{\partial l_\textrm{p}}{\partial \tpost{k}} + l_V^- -l_V^+\bigg]
        \end{aligned}$ \\
    \end{tabular}
    \caption{The adjoint spiking network to \cref{fig:lif} that computes the adjoint variable $\lambda_I$ needed for the gradient (\cref{eq:loss-gradient}).
        The adjoint variables are computed in reverse time (i.e., from $t=T$ to $t=0$) with $'=-\ddt$ denoting the reverse time derivative.
        $(\lambda_V^-)_{n(k)} $ experiences jumps at the spikes times $\tpost{k}$, where $n(k)$ is the index of the neuron that caused the $k$th spike.
        Computing this system amounts to the backpropagation of errors in time.
        The initial conditions are $\lambda_V(T)=\lambda_I(T)=0$ and we provide $\lambda_V^-$ in terms of $\lambda_V^+$ because the computation happens in reverse time.}\label{fig:adjoint-system}
\end{table}

\Cref{eq:loss-gradient,fig:adjoint-system} suggest a simple algorithm, EventProp, to compute the gradient (\cref{alg:eventprop}).
Notably, if the loss is voltage-independent (i.e., $l_V=0$), the backward pass of the algorithm requires only the spike times $\tpost{}$ and the synaptic current of the firing neurons at their respective firing times to be retained from the forward pass.
The membrane potential at spike times is fixed to the threshold $\vartheta$ and therefore implicitly retained; the synaptic current therefore determines the temporal derivative of the membrane potential at the spike time, $\dot V^-$, and needs to be stored for the backward pass.
The memory requirement of the algorithm scales as $\mathcal{O}(S)$, where $S$ is the number of post-synaptic spikes in the network.
A feed-forward architecture corresponds to a block matrix $W$ with each block being a strictly triangular matrix that connects two given layers.
In that case, the forward and backward pass can be computed in a layer-wise fashion.

In case of a voltage-dependent loss $l_V\neq 0$, the algorithm has to store the non-zero components of $\frac{\partial l_V}{\partial V}$ along the forward trajectory.
The loss $l_V$ may depend on the voltage at a discrete time $t_i$ using the Dirac delta, $l_V(V(t), t) = V(t)\delta(t_i-t)$, causing a jump of $\lambda_V$ of magnitude $\taum^{-1}$ at time $t_i$.
Note that in many practical scenarios as found in deep learning, the loss $l_V$ depends only on the state of a constant number of neurons, irrespective of network size.
If $l_V$ depends on the voltage of non-firing readout neurons, we have $l_V^+ = l_V^-$ and the corresponding term in the jump given in \cref{fig:adjoint-system} vanishes.

\begin{figure}[h!]
    \centering
    \includegraphics[width=\textwidth]{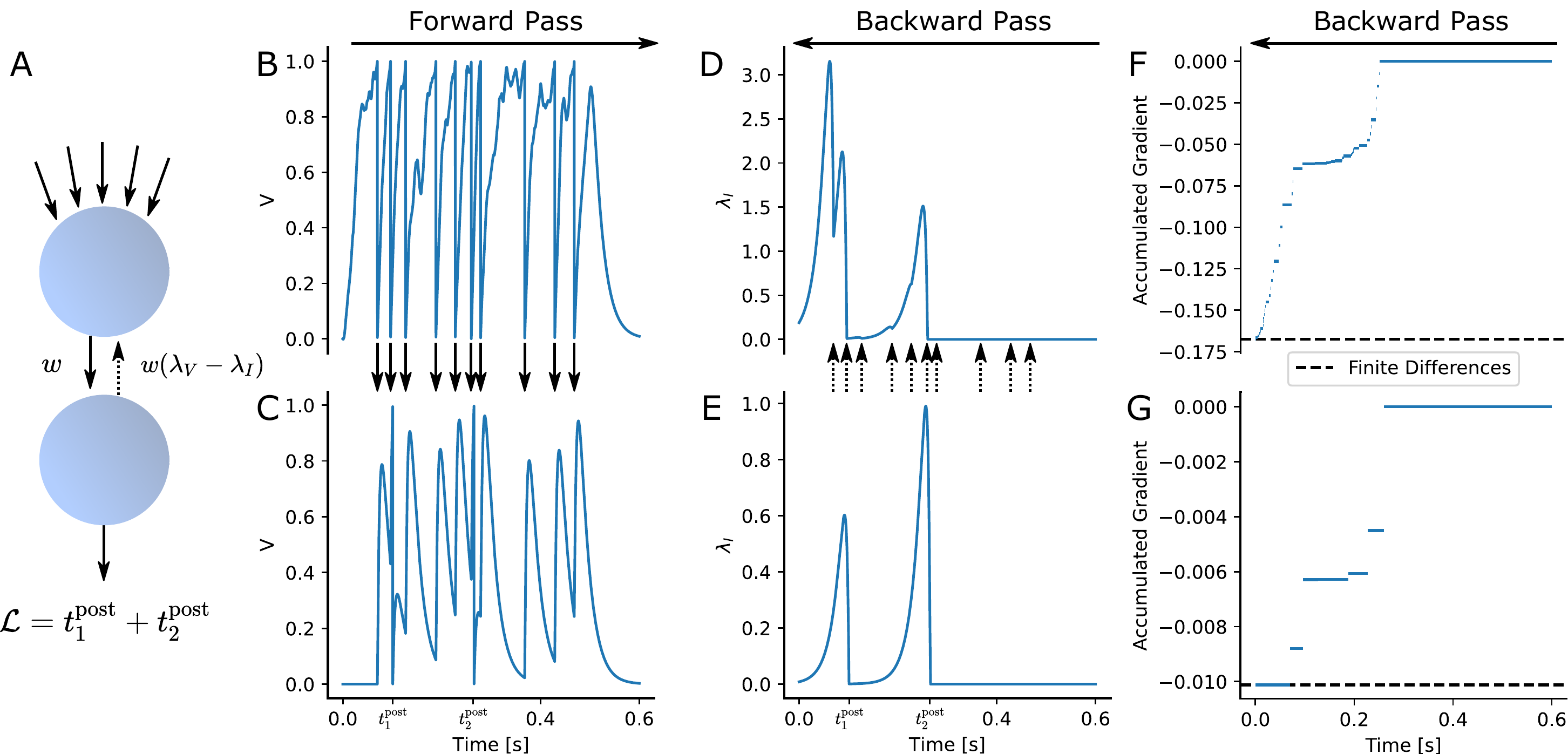}
    \caption{Illustration of EventProp-based gradient calculation in two leaky integrate-and-fire neurons connected with weight $w$ and a spike-time dependent loss $\mathcal{L}$.
        The forward pass (B, C) computes the spike times for both neurons and the backward pass (D-G) backpropagates errors at spike times, yielding the gradient as given in \cref{eq:loss-gradient}.
        \textbf{A}: The upper neuron receives $100$ independent Poisson spike trains with frequency \SI{200}{\hertz} across randomly initialized weights and is connected to the lower neuron via a single weight $w$.
        The loss $\mathcal{L}$ is a sum of the spike times of the lower neuron.
        \textbf{B, C}: Membrane potential of upper and lower neuron.
        Spike times of the upper neuron are indicated using arrows.
        \textbf{D, E}: Adjoint variable $\lambda_I$ of upper and lower neuron.
        The lower neuron backpropagates its error signal $\lambda_V-\lambda_I$ at the upper neuron's spike times (indicated by arrows).
        \textbf{F, G}: Accumulated gradient for one of the $100$ input weights of the upper neuron and the weight $w$ connecting the upper and lower neuron.
        EventProp computes the adjoint variables from $t=T$ to $t=0$ and accumulates the gradients by sampling $-\taus \lambda_I$ when spikes are transmitted across the respective weight.
        The gradients computed in this way match the gradients computed via central differences (dashed lines) up to a relative deviation of less than $10^{-7}$. }\label{fig:adjoint}
\end{figure}

If $l_V$ is either zero or depends only on voltages at discrete points in time, EventProp can be computed in a purely event-based manner.
\Cref{fig:adjoint} illustrates how EventProp computes the gradient of a spike time based loss function for two leaky integrate-and-fire neurons where one neuron receives Poisson spike trains via $100$ synapses and is connected to the other neuron via a single feed-forward weight $w$.

\begin{algorithm}[h!]
    \caption{EventProp: Algorithm to compute \cref{eq:loss-gradient}.}\label{alg:eventprop}
    \begin{algorithmic}[0]
        \Require Input spikes, losses $l_\mathrm{p}$, $l_V$, parameters $W$, $\taum$, $\taus$, initial conditions $V(0)$, $I(0)$
        \State $\mathrm{grad} \gets 0$
        \State    Compute neuron state trajectory (\cref{fig:lif}) from $t=0$ to $t=T$:
        \Comment Forward pass
        \State \hskip1.0em for all spikes $k$, store spike time $\tpost{k}$ and the firing neuron's component of $I(\tpost{k})$
        \State \hskip1.0em if $l_V\neq 0$, also store $\frac{\partial l_V}{\partial V}$
        \State Compute adjoint  state trajectory (\cref{fig:adjoint-system}) from $t=T$ to $t=0$:
        \Comment Backward pass
        \State \hskip1.0em accumulate
        \State \hskip1.5em $\mathrm{grad}_{ji}\gets\mathrm{grad}_{ji}-\taus (\lambda_I)_j$
        \State \hskip1.0em for each spike from neuron $i$ to $j$
        \State \Return grad
    \end{algorithmic}
\end{algorithm}

\clearpage
\subsection{Simulation Results}\label{sec:results}
We demonstrate learning using EventProp using a custom event-based simulator and the Yin-Yang \cite{Kriener2020} and MNIST \cite{LeCunMNIST} datasets.
In both cases, we use a single hidden layer and spike latency encoding of the input data.
The Yin-Yang dataset is classified using the time to first spike of a layer of readout neurons while the MNIST dataset is classified using the voltage maxima of a layer of non-firing readout neurons.
The simulator computes gradients using EventProp as described in \cref{alg:eventprop}; specifically, it uses an event queue and root-bracketing to compute post-synaptic spike times in the forward pass (using exact  integration of the membrane potential, \cite{Rotter1999}) and backpropagates errors by attaching error signals to spikes in the backward pass and using reverse traversal of the event queue.
We optimized synaptic weights using the calculated gradients via the Adam optimizer \citep{kingma2014adam}, without clipping gradients.

By initializing synaptic weights such that the network started in a non-quiescent state, we found that no explicit regularization of firing rates was needed to obtain the reported results in both cases.
Hyperparameters were optimized using Gaussian process optimization \citep{scikit-learn} and manual tuning using the validation set of the respective dataset.
The resulting parameters (see \cref{tab:sim-parameters}) were then evaluated using the test set.
\subsubsection{Yin-Yang Dataset}
\begin{figure}[h!]
    \centering
    \includegraphics[width=.9\textwidth]{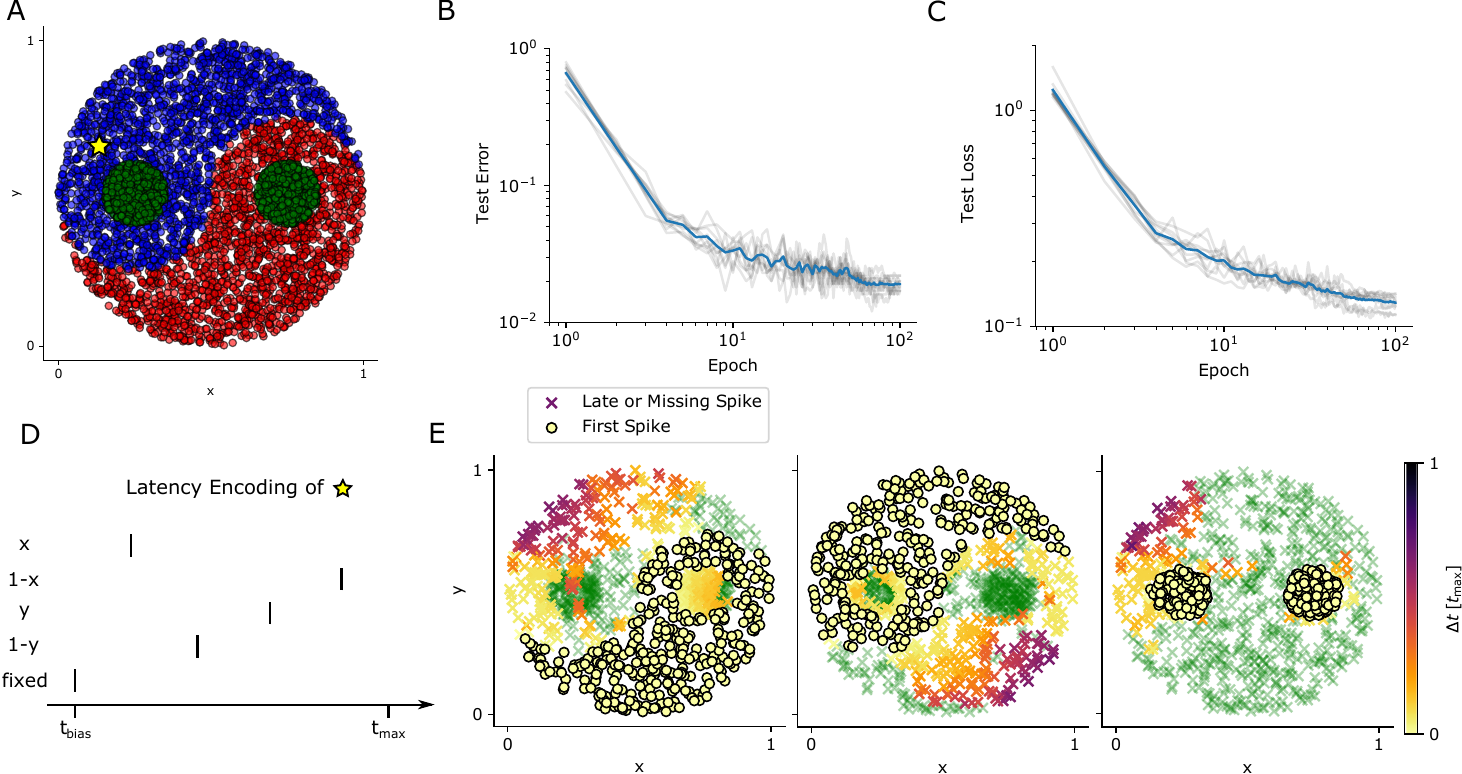}
    \caption{We used EventProp and a time-to-first-spike loss function to train a two-layer leaky integrate-and-fire network on the Yin-Yang dataset.
        \textbf{A}: Illustration of the two-dimensional training dataset.
        The three different classes are shown in red, green and blue.
        This dataset was encoded using spike time latencies (see D).
        \textbf{B, C}: Training results in terms of test error and loss averaged over $10$ different random seeds (individual traces shown as grey lines).
        \textbf{D}: Data points $(x, y)$ were transformed into $(x, 1-x, y, 1-y)$ and encoded using spike time latencies.
        We added a fixed spike at time $t_\textrm{bias}$.
        \textbf{E}: Spike time latencies $\Delta t$ of the three output neurons (encoding the blue, red or green class) after training, for all samples in the test set and a specific random seed.
        Latencies are relative to the first spike among the three neurons and given in units of $t_\mathrm{max}$.
        A latency of zero (bright yellow dots) implies that the corresponding neuron fired the first spike, determining the class assignment.
        Missing spikes are denoted using green crosses.}\label{fig:yinyang}
\end{figure}
The Yin-Yang dataset \cite{Kriener2020} is a two-dimensional non-linearly separable dataset, with a shallow classifier achieving around $64\%$ accuracy, and it therefore requires a hidden layer and backpropagation of errors for high classification accuracy.
Consider that in contrast, the MNIST dataset can be classified using a linear classifier with at least $88\%$ accuracy \citep{LeCunMNIST}.

Each two-dimensional data point of the dataset $(x,y)$ was transformed into four dimensions as $(x, 1-x, y, 1-y)$ and encoded using spike latencies in the interval $[0, t_\textrm{max}]$ (see \cref{fig:yinyang} D).
We added a fixed bias spike at time $t_\textrm{bias}$ for a total of five input spikes per data point.
The resulting spike patterns were used as input to a two-layer network composed of leaky integrate-and-fire neurons.
The output layer consisted of three neurons that each encoded one of the three classes, with each data point being assigned the class of the neuron that fired the earliest spike.

In analogy to \cite{gltz2019fast}, we used a cross-entropy loss defined using the first output spike times per neuron,
\begin{align}\label{eq:ttfs}
    \mathcal{L} = -\frac{1}{N_\mathrm{batch}}\left[\sum_{i=1}^{N_\mathrm{batch}}\log\left[\frac{\exp\left(-\tpost{i,l(i)}/\tau_0\right)}{\sum_{k=1}^3 \exp\left(-\tpost{i,k}/\tau_0\right)}\right] + \alpha \left[\exp\left(\frac{\tpost{i,l(i)}}{\tau_1}\right) - 1\right]\right],
\end{align}
where $\tpost{i,k}$ is the first spike time of neuron $k$ for the $i$th sample, $l(i)$ is the index of the correct label for the $i$th sample, $N_\mathrm{batch}$ is the number of samples in a given batch and $\tau_0$ and $\tau_1$ are hyperparameters of the loss function.
The first term corresponds to a cross-entropy loss function over the softmax function applied to the negative spike times (we use negative spike times as the class assignment is determined by the smallest spike time) and encourages an increase of the spike time difference between the label neuron and all other neurons.
As the first term depends only on the relative spike times, the second term is a regularization term that encourages early spiking of the label neuron.

Training results are shown in \cref{fig:yinyang}.
After training, the test accuracy was \SI{98.1(2)}{\percent} (mean and standard deviation over $10$ different random seeds).
This is comparable to the results shown in \cite{gltz2019fast}, who report \SI{95.9(7)}{\percent} accuracy with a smaller hidden layer ($200$ vs. $120$ neurons).

\subsubsection{MNIST Dataset}
\begin{figure}[h!]
    \centering
    \includegraphics[width=.9\textwidth]{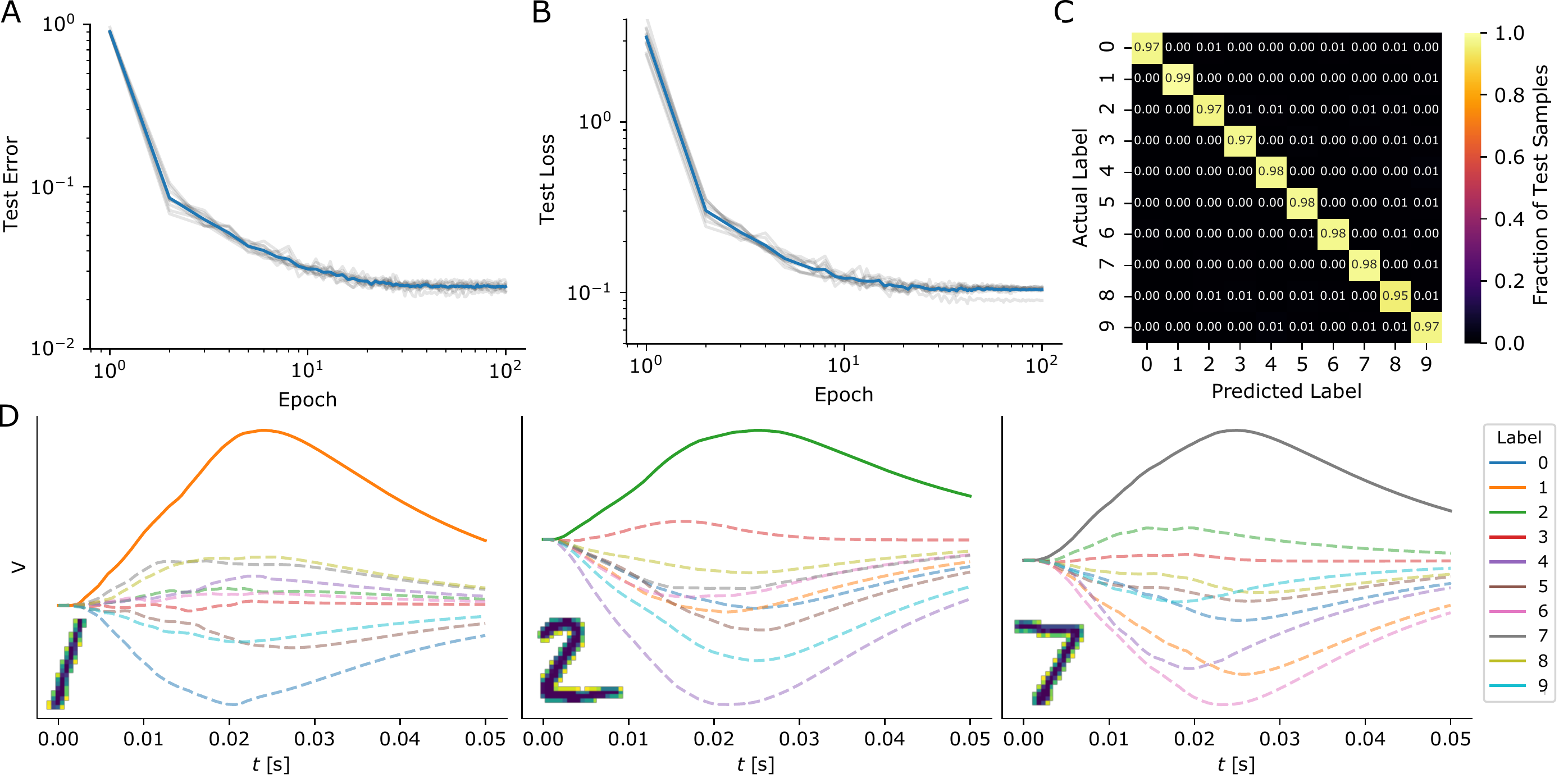}
    \caption{We used EventProp and a two-layer network composed of a hidden layer of leaky integrate-and-fire neurons and a readout layer of non-firing neurons to classify the MNIST dataset, with the readout neuron with the largest voltage deflection determining the class assignment.
    \textbf{A, B}: Training results in terms of test error and loss averaged over $10$ different random seeds (individual traces shown as grey lines).
    \textbf{C}: Confusion matrix after training for a specific random seed and using the test set.
    \textbf{D}: Voltage traces of all readout layer neurons for three different samples from the test set, where voltage traces of neurons corresponding to wrong labels are plotted using dashed lines.
    }\label{fig:mnist}
\end{figure}
We encoded each digit of the MNIST dataset \cite{LeCunMNIST} by transforming each of the $28\cdot 28=784$ pixels into spike latencies in the interval $[0, t_\mathrm{max}]$ (pixels corresponding to a value of $0$ or $1$ out of $255$ were not converted to spikes).
The resulting spike patterns were used as input to a two-layer network composed of a hidden layer of leaky integrate-and-fire neurons and a readout layer of non-firing leaky integrator neurons.
We used a cross-entropy loss function over the softmax function applied to the voltage maxima of the readout neurons (max-over-time),
\begin{align}
    \mathcal{L} = -\frac{1}{N_\mathrm{batch}}\sum_{i=1}^{N_\mathrm{batch}}\log\left[\frac{\exp\left(\max_t V_{l(i)}(t)\right)}{\sum_{k=1}^{10} \exp\left(\max_t V_k(t)\right)}\right],
\end{align}
where $V_k(t)$ is the voltage trace of the $k$th readout neuron, $l(i)$ is the index of the correct label for the $i$th sample and $N_\mathrm{batch}$ is the number of samples in a given batch.
Note that we can write the maximum voltage as $\max_t V_k(t)=\int V_k(t)\delta(t-t_\mathrm{max})\mathrm{d}t$ with the time of the maximum $t_\mathrm{max}$ and the Dirac delta $\delta$, allowing us to apply the chain rule to find the jump of $\lambda_{V_k}$ (cf. \cref{fig:adjoint-system}) at time $t_\mathrm{max}$ (terms containing the distributional derivative of $\delta$ are always zero).

During training, input spikes were dropped with probability $p_\mathrm{drop}$ in order to avoid overfitting.
To obtain a validation set, we extracted and removed $5000$ samples from the training set.

Training results are shown in \cref{fig:mnist}.
After training, the test accuracy was \SI{97.6(1)}{\percent} (mean and standard deviation over $10$ different random seeds).
This represents competitive classification performance when compared with previously published results using spiking networks with a single, fully connected hidden layer (\cref{tab:mnist}).

\begin{table}[h!]
    \small{
    \begin{tabular}{llll}
        \textbf{Publication} & \textbf{\# Hidden} & \textbf{Test Accuracy}  & \textbf{Comments}  \\ \hline
        This Work & 350 & \SI{97.6(1)}{\percent}&\\
        Cramer et al., 2021 \cite{cramer2021surrogate} & 246 & \SI{97.5(1)}{\percent} & Downsampled to $16$ by $16$ pixels \\
        Zenke \& Vogels, 2021 \cite{zenke2021} & 100 & \SI{98.3(9)}{\percent} & \\
        Kheradpisheh \& Masquelier, 2020 \cite{Kheradpisheh2020} & 400 & \SI{97.4(2)}{\percent}& \\
        Comsa et al., 2020 \cite{Comsa2020} & 340 & \SI{97.9}{\percent} (Max.) & Bias spikes at learned times\\
        Göltz et al., 2019 \cite{gltz2019fast} & 350 & \SI{97.5(1)}{\percent}&\\
        Mostafa, 2017 \cite{Mostafa_2017} & 800 & \SI{97.55}{\percent} &\\
        Neftci et al., 2017 \cite{Neftci2017} & 500 & \SI{97.77}{\percent} (Max.) &\\
        Lee et al., 2016 \cite{lee2016} & 800 & \SI{98.71}{\percent} (Max.) &
    \end{tabular}
    \caption{Comparison of previously published classification results on the MNIST dataset for spiking neural networks that are trained using supervised learning with a single, fully connected (non-convolutional) hidden layer and temporal encoding of input data.
    The second column provides the number of hidden neurons.}\label{tab:mnist}}
\end{table}

\section{Discussion}
We have derived and provided an algorithm (EventProp) to compute the gradient of a general loss function for a spiking neural network composed of leaky integrate-and-fire neurons.
The parameter-dependent spike discontinuities were treated in a well-defined manner using the adjoint method in combination with partial derivative jumps, without approximations or smoothing operations.
EventProp uses the resulting adjoint spiking network to backpropagate errors in order to compute the exact gradient.
Its forward pass requires computing the spike times of pre-synaptic neurons that transmit spikes to post-synaptic neurons, while the backward pass backpropagates errors at these spike times using the reverse path (i.e., from post-synaptic to pre-synaptic neurons).
The rigorous treatment of spike discontinuities in combination with an event-based computation of the exact gradient represent a significant conceptual advance in the study of gradient-based learning methods for spiking neural networks.

An apparent issue with gradient descent based learning in the context of spiking networks is that the magnitude of the gradient diverges at the critical points in parameter space (note the $\dot v^{-1}$ term in the jump term given in \cref{fig:adjoint-system}; this term diverges as the membrane potential becomes tangent to the threshold and we have $\dot v\to 0$).
Indeed, this is a known issue in the broader context of optimal control of dynamical systems with parameter-dependent state transitions \citep{Barton2002,GALAN199917}.
While this divergence can be mitigated using gradient clipping in practice, exact gradients of commonly considered loss functions lead to learning dynamics that are ignorant with respect to these critical points and are therefore unable to selectively recruit additional spikes or dismiss existing spikes.
In contrast, surrogate gradient methods continuously transmit errors across neurons and combine these with a non-linear function of the distance of the membrane potential to the threshold.
It is therefore plausible that surrogate gradients represent a form of implicit regularization.
\cite{neftci:2019} reports that the surrogate gradient approximates the true gradient in a minimalistic binary classification task while at the same time remaining finite and continuous along an interpolation path in weight space.
Hybrid algorithms that combine the exact gradient with explicit regularization techniques could be a direction for future research and provide more principled learning algorithms as compared to ad-hoc replacements of threshold functions.

This work is based on the widely used leaky integrate-and-fire neuron model.
Extensions to this model, such as fixed refractory periods, adaptive thresholds or multiple compartments can be treated in an analogous way \cite{pehle2021adjoint}.
While the absence of explicit solutions to the resulting differential equations can require the use of sophisticated numerical techniques for event-based simulations, such extensions can significantly enhance the computational capabilities of spiking networks.
For example, \cite{bellec:2018} uses adaptive thresholds to implement LSTM-like memory cells in a recurrent spiking neural network.

Neuromorphic hardware is an increasingly active research subject \citep[e.g., ][]{aamir:2018,loihi:2018,spinnaker:2014,Neckar2019,Moradi2018,merolla2014million,Pei2019,billaudelle2019versatile,feldmann2019,Boybat2017,Wunderlich_2019} and implementing EventProp on such hardware is a natural consideration.
The adjoint dynamics as given in \cref{fig:adjoint-system} represent a type of spiking neural network which, instead of spiking dynamically, transmits errors at fixed times $\tpost{}$ that are scaled with factors $\dot v^{-1}$ retained from the forward pass.
Therefore, a neuromorphic implementation could store spike times and scaling factors locally at each neuron, where they could be combined with the dynamic error signal ($\lambda_V -\lambda_I$ in \cref{fig:adjoint-system}) in the backward pass.
This requires a possibility to read out neuronal state variables both in the forward and backward pass (membrane potential and synaptic current).
The resulting error signals could be distributed across the network using event-based communication schemes similar to, for example, the address-event representation protocol \citep{Chan2007}.
As mentioned above, EventProp can be extended to multi-compartment neuron models as used in a recent neuromorphic architecture \citep{schemmel2017accelerated}.

We used a two-layer feed-forward architecture to demonstrate learning using EventProp.
The algorithm can, however, compute the gradient for arbitrary recurrent or convolutional architectures.
Its computational and spatial complexity scales linearly with network size (assuming constant average firing rates per neuron), analogous to backpropagation in non-spiking artificial neural networks.
The performance in more complex tasks therefore hinges on the general efficacy of gradient-based optimization in spiking networks.
As mentioned above, gradients with respect to loss functions defined in terms of spike times or membrane potentials ignores the presence of critical parameters where spikes appear or disappear.
We suggest that studying regularization techniques which deal with this fundamental issue  in a targeted manner could enable powerful learning algorithms for spiking networks.
By providing a theoretical foundation for backpropagation in spiking networks, we support future research that combines such regularization techniques with the computation of exact gradients.

\clearpage
\input{tex/derivation}

\section*{Contributions}
CP conceived of the presented idea and outlined the application of sensitivity analysis
to spiking neuron models. CP and TW derived the adjoint equations for LIF
neurons and the resulting EventProp algorithm. TW implemented the event based simulation code.
TW conducted and analyzed the presented simulations.
CP and TW wrote and edited the manuscript.
\section*{Acknowledgements}
We would like to express gratitude to Eric Mülller and Johannes Schemmel for discussions, continued
 support and encouragement during the preparation of this work. We thank
 Laura Kriener and Julian Göltz for their support
 regarding time to first spike experiments and helpful discussions. We thank Korbinian Schreiber and Mihai Petrovici for helpful discussions.
\section*{Funding}
This research has received funding from the EC Horizon 2020 Framework Programme under Grant Agreements 785907 and 945539 (HBP) and was financially supported by the Joachim Herz foundation.

\bibliography{references}
\clearpage
\appendix
\section{Simulation Parameters}
\begin{table}[h!]
    \begin{tabular}{llll}
        \textbf{Symbol}    & \textbf{Description}                & \textbf{Value (Yin-Yang Dataset)} & \textbf{Value (MNIST Dataset)}   \\ \hline
        $\taum$            & Membrane time constant              & $\SI{20}{\milli\second}$ & $\SI{20}{\milli\second}$ \\
        $\taus$            & Synaptic time constant              & $\SI{5}{\milli\second}$  & $\SI{5}{\milli\second}$  \\
        $\vartheta$        & Threshold                           & $1$                      & $1$                      \\
                           & Input size                          & $5$                      & $784$                    \\
                           & Hidden size                         & $200$                    & $350$                    \\
                           & Output size                         & $3$                      & $10$                     \\
        $t_\mathrm{bias}$  & Bias time                           & \SI{0}{\milli\second}    & n/a                      \\
        $t_\mathrm{max}$   & Maximum time                        & \SI{30}{\milli\second}   & \SI{20}{\milli\second}   \\
                           & Hidden weights mean                 & $1.5$                    & $0.078$                  \\
                           & Hidden weights standard deviation   & $0.78$                   & $0.045$                  \\
                           & Output weights mean                 & $0.93$                   & $0.2$                    \\
                           & Output weights standard deviation   & $0.1$                    & $0.37$                   \\
                           & Minibatch size                      & $32$                    & $5$                      \\
                           & Optimizer                           & Adam                     & Adam                     \\
        $\beta_1$          & Adam parameter                      & $0.9$                    & $0.9$                    \\
        $\beta_2$          & Adam parameter                      & $0.999$                  & $0.999$                  \\
        $\epsilon$         & Adam parameter                      & \num{1e-8}               & \num{1e-8}               \\
        $\eta$             & Learning rate                       & \num{5e-3}               & \num{5e-3}               \\
                           & Learning rate decay factor          & $0.95$                   & $0.95$                   \\
                           & Learning rate decay step            & $1$ epoch               & $1$ epoch                \\
        $p_\mathrm{drop}$  & Prob. of dropping input spike       & n/a                      & 0.2                      \\
        $\alpha$           & Regularization factor               & \num{3e-3}               & n/a                      \\
        $\tau_0$           & First loss time constant            & \SI{0.5}{\milli\second}  & n/a                      \\
        $\tau_1$           & Second loss time constant           & \SI{6.4}{\milli\second}  & n/a
    \end{tabular}
    \caption{Simulation parameters used for the results shown in \cref{sec:results}.} \label{tab:sim-parameters}
\end{table}

\end{document}

%% file: tex/derivation.tex
\section{Methods}
\subsection{Partial Derivatives in a Hybrid System}
\label{sec:pdh}
In the following, we use the example of a bouncing ball (\Cref{fig:fig1} A) to illustrate the calculation of partial derivatives in a dynamical system with state discontinuities.
A general treatment of the topic is given in \cite{GALAN199917} or \cite{barton1998}.
The discontinuities occurring in the leaky integrate-and-fire neuron are  treated analogously in our derivation of the gradient (\cref{sec:control-derivation}).

The differential equation describing the bouncing ball with height $y$ is
\begin{align}
	\ddot y = -g
\end{align}
with gravitational acceleration $g$.
Defining the ball's velocity as $v\equiv \dot y$, this is equivalent to a two-dimensional system
\begin{subequations}
\begin{align}
	\dot v &= - g,\\
	\dot y &= v.
\end{align} \label{eqs:ball}
\end{subequations}
The initial conditions are
\begin{subequations}
\begin{align}
	v(0) &= 0, \\
	y(0) &= y_0
\end{align}\label{eqs:ballic}
\end{subequations}
where $y_0>0$ is the parameter of interest defining the ball's initial height.
The given equations determine the state trajectory $y(t)$ up to the moment of impact with the ground at $y=0$.
Likewise, the trajectories of the partial derivatives with respect to $y_0$ are given by differentiation of \cref{eqs:ball,eqs:ballic} \cite{Gronwall1919},
\begin{subequations}
	\begin{align}
		\frac{\mathrm{d}}{\mathrm{d}t}\frac{\partial v}{\partial y_0} &= 0,\\
		\frac{\mathrm{d}}{\mathrm{d}t}\frac{\partial y}{\partial y_0} &= \frac{\partial v}{\partial y_0},
	\end{align}
\end{subequations}
with initial conditions
\begin{subequations}
	\begin{align}
        \frac{\partial v}{\partial y_0}(0) &= 0,\\
		\frac{\partial y}{\partial y_0}(0) &= 1.
	\end{align}
\end{subequations}
The state discontinuity occurs when the ball hits the ground and we have
\begin{align}
	y^- = 0 \label{eq:balltranscond}
\end{align}
at the time of impact $t_\mathrm{r}$.
The ball is inelastically reflected, losing a fraction of its energy.
Specifically, the system is re-initialized as
\begin{subequations}
	\begin{align}
		v^+ &= -0.8v^-,\\
		y^+ &= y^-,
	\end{align}\label{eqs:ballreinit}
\end{subequations}
where $-$ and $+$ denote the state before and after the transition ($v^\pm$, $y^\pm$ are functions of $t_r$ and $y_0$).
\Cref{eq:balltranscond} and \cref{eqs:ballreinit} together uniquely determine the partial derivatives after the reflection.
The implicit function theorem \cite{krantz2012implicit} applied to \cref{eq:balltranscond} guarantees (because $v\neq 0$) the existence of a function $t_\mathrm{r}(y_0)$ that locally describes how the impact time changes with $y_0$, with its derivative given by
\begin{align}
	\frac{\mathrm{d}t_\mathrm{r}}{\mathrm{d}y_0} =-\frac{1}{\dot y^-}\frac{\partial y^-}{\partial y_0} = -\frac{1}{v^-}\frac{\partial y^-}{\partial y_0}. \label{eq:ball-imf}
\end{align}
Likewise, the implicit function theorem applies to \cref{eqs:ballreinit} (because $v\neq 0$, $\dot v \neq 0$), yielding after differentiation
\begin{subequations}
\begin{align}
	\frac{\partial v^+}{\partial y_0} + \dot v^+\frac{\mathrm{d}t_r}{\mathrm{d}y_0} &= \frac{\partial v^-}{\partial y_0} + \dot v^-\frac{\mathrm{d}t_r}{\mathrm{d}y_0},\\
	\frac{\partial y^+}{\partial y_0} + \dot y^+\frac{\mathrm{d}t_r}{\mathrm{d}y_0} &= \frac{\partial y^-}{\partial y_0} + \dot y^-\frac{\mathrm{d}t_r}{\mathrm{d}y_0}.
\end{align}\label{eqs:pdreinit}
\end{subequations}

The partial derivatives after the transition can now be found by solving the system of equations given by \cref{eqs:ballreinit,eq:ball-imf,eqs:pdreinit},
\begin{subequations}
\begin{align}
	\frac{\partial v^+}{\partial y_0} &= -0.8\frac{\partial v^-}{\partial y_0}-1.8g\frac{1}{v^-}\frac{\partial y^-}{\partial y_0},\\
	\frac{\partial y^+}{\partial y_0} &= -0.8\frac{\partial y^-}{\partial y_0},
\end{align}\label{eqs:pdreinitsolved}
\end{subequations}
where we have used $\ddot y = -g$.
\Cref{eqs:pdreinitsolved} provides the initial conditions for the integration of the partial derivatives after the transition; subsequent ground impacts can be treated equivalently.
\Cref{fig:fig1} A illustrates the behaviour of $y(t)$ and $\frac{\partial y}{\partial y_0}(t)$ using trajectories calculated numerically using the equations given here.

\subsection{Adjoint Method }\label{sec:adjoint}
We apply the adjoint method to a continuous, first order system of ordinary differential equations and refer the reader to \cite{pontryagin:1962,bradley:2019} for a more general setting.
Consider an $N$-dimensional dynamical system $x: t\mapsto x(t)\in\mathbb{R}^N$ with parameters $p\in \mathbb{R}^P$ defined by the  system of implicit first order  ordinary differential equations
\begin{align} \label{eq:adjoint-method-ode}
	\dot x - F(x, p) = 0
\end{align}
and constant initial conditions $G(x(0))=0$ where $F$, $G$ are smooth vector-valued functions.

We are interested in computing the gradient of a loss that is the integral of a smooth function $l$ over the trajectory of $x$,
\begin{align}
	\mathcal L = \int_0^T l(x, t)\dt.
\end{align}
We have
\begin{align} \label{eq:adjoint-method-gradient}
	\frac{\mathrm{d}\mathcal{L}}{\mathrm{d}p_i} = \int_0^T \frac{\partial l}{\partial x}\cdot\frac{\partial x}{\partial p_i} \dt,
\end{align}
where $\cdot$ is the dot product and the dynamics of the partial derivatives $\frac{\partial x}{\partial p_i}$ are given by applying Gronwall's theorem \citep{Gronwall1919},
\begin{align}\label{eq:adjoint-method-partial-ode}
	\ddt \frac{\partial x}{\partial p_i} = \frac{\partial F}{\partial x}\frac{\partial x}{\partial p_i} + \frac{\partial F}{\partial p_i}.
\end{align}
Computing $x(t)$ along with $\frac{\partial x}{\partial p_i}(t)$ using \cref{eq:adjoint-method-ode,eq:adjoint-method-partial-ode}  allows us to calculate the gradient in \cref{eq:adjoint-method-gradient} in a single forward pass.
However, this procedure can incur prohibitive computational cost.
When considering a recurrent neural network with $N$ neurons and $P=N^2$ synaptic weights, computing  $\frac{\partial x}{\partial p_i}(t)$ for all parameters requires storing and integrating $PN=N^3$ partial derivatives.

The adjoint method allows us to avoid computing $PN$ partial derivatives in the forward pass by instead computing $N$ adjoint variables $\lambda(t)$ in an additional backward pass.
We add a Lagrange multiplier $\lambda: t\mapsto \lambda(t)\in\mathbb{R}^N$ that constrains the system dynamics as given in \cref{eq:adjoint-method-ode},
\begin{align}
	\mathcal L = \int_0^T\left[ l(x, t)+\lambda \cdot \left(\dot x - F(x,p)\right)\right]\dt.
\end{align}
Along trajectories where \cref{eq:adjoint-method-ode} holds, $\lambda$ can be chosen arbitrarily without changing $\mathcal{L}$ or its derivative.
We get
\begin{align} \label{eq:adjoint-method-adjoint-gradient}
	\frac{\mathrm{d}\mathcal{L}}{\mathrm{d}p_i} = \int_0^T \left[\frac{\partial l}{\partial x} \cdot \frac{\partial x}{\partial p_i} +\lambda  \cdot \left(\ddt \frac{\partial x}{\partial p_i} - \frac{\partial F}{\partial x}\frac{\partial x}{\partial p_i}- \frac{\partial F}{\partial p_i}\right)\right]\dt.
\end{align}
Using partial integration, we have
\begin{align}
	\int_0^T\lambda \cdot  \ddt \frac{\partial x}{\partial p_i}\dt = -\int_0^T\dot \lambda \cdot  \frac{\partial x}{\partial p_i}\dt + \left[\lambda\cdot \frac{\partial x}{\partial p_i} \right]_0^T.
\end{align}
By setting $\lambda(T)=0$, the boundary term vanishes because we chose parameter independent initial conditions  ($\frac{\partial x}{\partial p_i}(0)=0$).
The gradient becomes
\begin{align}
	\frac{\mathrm{d}\mathcal{L}}{\mathrm{d}p_i} = \int_0^T \left[ \left(\frac{\partial l}{\partial x} - \dot \lambda - \frac{\partial F}{\partial x}\lambda\right)\cdot \frac{\partial x}{\partial p_i} - \lambda\cdot\frac{\partial F}{\partial p_i}\right]\dt.
\end{align}
By choosing $\lambda$ to fulfill the adjoint differential equation
\begin{align}\label{eq:adjoint-ode}
	\dot \lambda = \frac{\partial l}{\partial x} - \frac{\partial F}{\partial x}\lambda
\end{align}
we are left with
\begin{align}\label{eq:adjoint-cont-grad}
	\frac{\mathrm{d}\mathcal{L}}{\mathrm{d}p_i} = - \int_0^T\lambda\cdot\frac{\partial F}{\partial p_i}\dt .
\end{align}
The gradient can therefore be computed using \cref{eq:adjoint-cont-grad}, where the adjoint state variable $\lambda$ is computed from $t=T$ to $t=0$ as the solution of the adjoint differential equation \cref{eq:adjoint-ode} with initial condition $\lambda(T)=0$.
This corresponds to backpropagation through time (BPTT) in discrete time artificial neural networks.

\subsection{Derivation of Gradient}\label{sec:control-derivation}
We apply the adjoint method (\cref{sec:adjoint}) to the case of a spiking neural network (i.e., a hybrid, discontinuous system with parameter dependent state transitions).
The following derivation is specific to the model given in \cref{fig:lif}.
A fully general treatment of (adjoint) sensitivity analysis in hybrid systems can be found in \cite{GALAN199917} or \cite{Serban_2019}.

The differential equations defining the free dynamics in implicit form are
\begin{subequations} \label{eqs:implicit-lif}
	\begin{align}
		f_V & \equiv \taum \dot V +V -I = 0, \\
		f_I & \equiv \taus \dot I +I = 0,
	\end{align}
\end{subequations}
where $f_V$, $f_I$ are again vectors of size $N$.
We now split up the loss integral in \cref{eq:loss-basic} at the spike times $\tpost{}$ and use vectors of Lagrange multipliers $\lambda_V$, $\lambda_I$ that fix the system dynamics $f_V$, $f_I$ between transitions.
\begin{align}
	\frac{\mathrm{d}\mathcal{L}}{\mathrm{d}w_{ji}} = \frac{\mathrm{d}}{\mathrm{d}w_{ji}}\left[l_\mathrm{p}(\tpost{})+\sum_{k=0}^{\npost} \int_{\tpost{k}}^{\tpost{k+1}} \left[ l_V(V,t) + \lambda_V\cdot f_V + \lambda_I\cdot f_I\right]\dt\right],
\end{align}
where we set $\tpost{0}=0$ and $\tpost{\npost+1}=T$ and $x\cdot y$ is the dot product of two vectors $x$, $y$.
Note that because $f_V$, $f_I$ vanish along all considered trajectories, $\lambda_V$ and $\lambda_I$ can be chosen arbitrarily without changing $\mathcal{L}$ or its derivative.
Using \cref{eqs:implicit-lif} we have, as per Gronwall's theorem \citep{Gronwall1919},
\begin{subequations}
	\begin{align}
		\frac{\partial f_V}{\partial w_{ji}} & = \taum \ddt {\frac{\partial V}{\partial w_{ji}}} + \frac{\partial V}{\partial w_{ji}} - \frac{\partial I}{\partial w_{ji}}, \\
		\frac{\partial f_I}{\partial w_{ji}} & = \taus \ddt {\frac{\partial I}{\partial w_{ji}}} + \frac{\partial I}{\partial w_{ji}},
	\end{align}
\end{subequations}
where we have used the fact that the derivatives commute, $\frac{\partial}{\partial w_{ji}} \frac{\mathrm{d}}{\mathrm{d}t} = \frac{\mathrm{d}}{\mathrm{d}t} \frac{\partial}{\partial w_{ji}}$ (the weights are fixed and have no time dependence).
The gradient then becomes, by application of the Leibniz integral rule,
\begin{align}
	\frac{\mathrm{d}\mathcal{L}}{\mathrm{d}w_{ji}} & = \sum_{k=0}^{\npost} \bigg[\int_{\tpost{k}}^{\tpost{k+1}}\left[ \frac{\partial l_V}{\partial V} \cdot \frac{\partial V}{\partial w_{ji}} + \lambda_V \cdot\left(\taum \ddt {\frac{\partial V}{\partial w_{ji}}} + \frac{\partial V}{\partial w_{ji}} - \frac{\partial I}{\partial w_{ji}}\right) + \lambda_I \cdot\left(\taus \ddt {\frac{\partial I}{\partial w_{ji}}} + \frac{\partial I}{\partial w_{ji}}\right)\right] \dt\nonumber        \\
	                                               & \quad+\frac{\partial l_\mathrm{p}}{\partial \tpost{k}}\frac{\mathrm{d}\tpost{k}}{\mathrm{d}w_{ji}}+l^-_{V,k+1}\frac{\mathrm{d}\tpost{k+1}}{\mathrm{d}w_{ji}} - l^+_{V,k}\frac{\mathrm{d}\tpost{k}}{\mathrm{d}w_{ji}}\bigg],
\end{align}
where $l_{V,k}^\pm$ is the voltage-dependent loss evaluated before ($-$) or after ($+$) the transition and we have used that $f_V=f_I=0$ along all considered trajectories.
Using partial integration, we have
\begin{align}
	\int_{\tpost{k}}^{\tpost{k+1}} \lambda_V \cdot \ddt \frac{\partial V}{\partial w_{ji}}\dt & = - \int_{\tpost{k}}^{\tpost{k+1}} \dot {\lambda}_V\cdot \frac{\partial V}{\partial w_{ji}}\dt + \bigg[ \lambda_V\cdot \frac{\partial V}{\partial w_{ji}} \bigg]_{\tpost{k}}^{\tpost{k+1}},  \\
	\int_{\tpost{k}}^{\tpost{k+1}} \lambda_I \cdot \ddt \frac{\partial I}{\partial w_{ji}}\dt & = - \int_{\tpost{k}}^{\tpost{k+1}} \dot {\lambda}_I\cdot \frac{\partial I}{\partial w_{ji}}\dt + \bigg[ \lambda_I \cdot \frac{\partial I}{\partial w_{ji}} \bigg]_{\tpost{k}}^{\tpost{k+1}}.
\end{align}

Collecting terms in $\frac{\partial V}{\partial w_{ji}}$, $\frac{\partial I}{\partial w_{ji}}$, we have
\begin{align}\label{eq:adj-med-deriv-int}
	\frac{\mathrm{d}\mathcal{L}}{\mathrm{d}w_{ji}} & =\sum_{k=0}^{\npost} \bigg[\int_{\tpost{k}}^{\tpost{k+1}}\bigg[\bigg( \frac{\partial l_V}{\partial V} - \taum \dot \lambda_V + \lambda_V\bigg) \cdot \frac{\partial V}{\partial w_{ji}} + \left(-\taus \dot \lambda_I + \lambda_I - \lambda_V\right)\cdot \frac{\partial I}{\partial w_{ji}} \bigg]\dt \nonumber      \\
	                                               & \quad+\frac{\partial l_\mathrm{p}}{\partial \tpost{k}}\taupost{k} +\taum\big[ \lambda_V \cdot \frac{\partial V}{\partial w_{ji}} \big]_{\tpost{k}}^{\tpost{k+1}}+\taus\big[ \lambda_I \cdot \frac{\partial I}{\partial w_{ji}} \big]_{\tpost{k}}^{\tpost{k+1}} +l^-_{V,k+1}\taupost{k+1} -l^+_{V,k}\taupost{k}\bigg].
\end{align}
Since the Lagrange multipliers $\lambda_V(t)$, $\lambda_I(t)$ can be chosen arbitrarily, this form allows us to set the dynamics of the adjoint variables between transitions.
Since the integration of the adjoint variables is done from $t=T$ to $t=0$ in practice (i.e., reverse in time), it is practical to transform the time derivative as $\frac{\mathrm{d}}{\mathrm{d}t}\to -\frac{\mathrm{d}}{\mathrm{d}t}$.
Denoting the new time derivative by $'$, we have
\begin{subequations} \label{eqs:adjoint-dynamics-derivation}
	\begin{align}
		\taum \lambda_V ' & = -\lambda_V - \frac{\partial l_V}{\partial V}, \\
		\taus \lambda_I ' & = -\lambda_I  + \lambda_V.
	\end{align}
\end{subequations}
The integrand in \cref{eq:adj-med-deriv-int} therefore vanishes along the trajectory and we are left with a sum over the transitions.
Since the initial conditions of $V$ and $I$ are assumed to be parameter independent, we have $\frac{\partial V}{\partial w_{ji}}=\frac{\partial I}{\partial w_{ji}}=0$ at $t=0$.
We set the initial condition for the adjoint variables to be $\lambda_V(T)=\lambda_I(T)=0$ to eliminate the boundary term for $t=T$.
We are therefore left with a sum over transitions $\xi_k$ evaluated at the transition times $\tpost{k}$,
\begin{align}
	\frac{\mathrm{d}\mathcal{L}}{\mathrm{d}w_{ji}} & =\sum_{k=1}^{\npost} \xi_{k}
\end{align}
with the definition
\begin{align}
	\xi_k & \equiv \frac{\partial l_\mathrm{p}}{\partial \tpost{k}}\taupost{k}+l_{V,k}^-\taupost{k} - l_{V,k}^+\taupost{k}\nonumber                                                                       \\
	      & \quad+\left[\taum\left(\lambda_{V}^-\cdot \frac{\partial V^-}{\partial w_{ji}} - \lambda_{V}^+\cdot \frac{\partial V^+}{\partial w_{ji}}\right) + \taus \left(\lambda_{I}^-\cdot \frac{\partial I^-}{\partial w_{ji}} - \lambda_{I}^+\cdot \frac{\partial I^+}{\partial w_{ji}}\right)\right]\bigg\rvert_{\tpost{k}}. \label{eq:xi-def}
\end{align}

We proceed by deriving the relationship between the adjoint variables before and after each transition.
Since the computation of the adjoint variables happens in reverse time in practice, we provide $\lambda^-$ in terms of $\lambda^+$.

Consider a spike caused by the $n$th neuron, with all other neurons $m\neq n$ remaining silent.
We start by first deriving the relationships between $\frac{\partial V^+}{\partial w_{ji}}$, $\frac{\partial V^-}{\partial w_{ji}}$ and $\frac{\partial I^+}{\partial w_{ji}}$, $\frac{\partial I^-}{\partial w_{ji}}$.
\paragraph{Membrane Potential Transition}
\begin{wrapfigure}{l}{.4\textwidth}
	\centering
	\includegraphics[width=.35\textwidth]{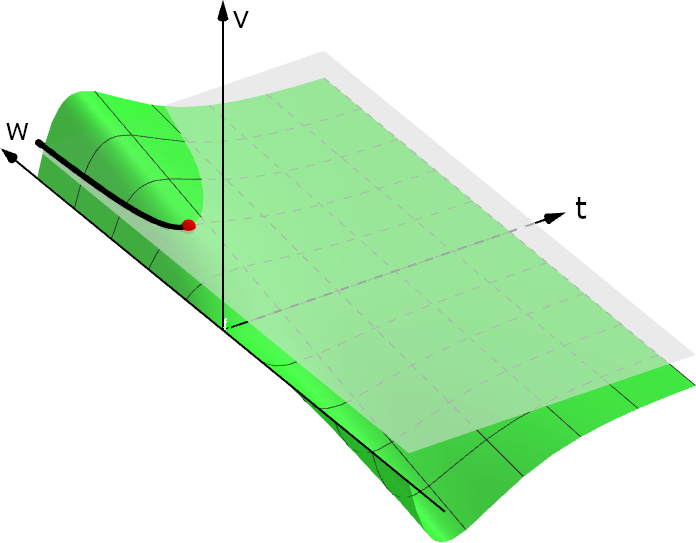}
	\caption{In this sketch, the relation $v(t,w)-\vartheta=0$ defines an implicit function (black line along which $\mathrm{d}v=0$).
		The critical point where the gradient diverges is shown in red.}\label{fig:impl-spike}
	\vspace{-.5cm}
\end{wrapfigure}

By considering the relations between $V^+$, $V^-$ and $\dot V^+$, $\dot V^-$, we can derive the relation between $\frac{\partial V^+}{\partial w_{ji}}$ and $\frac{\partial V^-}{\partial w_{ji}}$ at each spike.
Each spike at $\tpost{}$ is triggered by a neuron's membrane potential crossing the threshold.
We therefore have, at $\tpost{}$,
\begin{align}
	(V^-)_n - \vartheta = 0.
\end{align}
This relation defines $\tpost{}$ as a differentiable function of $w_{ji}$ via the implicit function theorem \citep[illustrated in \cref{fig:impl-spike}, see also][]{wenyu2014}, under the condition that $(\dot V^-)_n\neq 0$.
Differentiation of this relation yields
\begin{align} \label{eq:postdiff}
	\left(\frac{\partial V^-}{\partial w_{ji}}\right)_n + (\dot V^-)_n\taupost{} = 0.
\end{align}
Since we only allow transitions for $(\dot V^-)_n\neq 0$, we have
\begin{align}\label{eq:taupost}
	\taupost{} = -\frac{1}{(\dot V^-)_n} \left(\frac{\partial V^-}{\partial w_{ji}}\right)_n.
\end{align}
Note that corresponding relations were previously used to derive gradient-based learning rules for spiking neuron models \citep{NIPS2004_2674,bohte:2000spikeprop,Booij2005,Yan2013,Florian2012}; in contrast to the suggestion in \cite{bohte:2000spikeprop}, \cref{eq:taupost} is not an approximation but rather an exact relation at all non-critical parameters and invalid at all critical parameters.

Because the spiking neuron's membrane potential is reset to zero, we have
\begin{align}
	(V^+)_n = 0.
\end{align}
This implies by differentiation
\begin{align}
	\left(\frac{\partial V^+}{\partial w_{ji}}\right)_n + (\dot{V}^+)_n\taupost{} & = 0.
\end{align}
Using \cref{eq:taupost}, this allows us to relate the partial derivative after the spike to the partial derivative before the spike,
\begin{align}
	\left(\frac{\partial V^+}{\partial w_{ji}}\right)_n             & = \frac{(\dot V^+)_n}{(\dot V^-)_n} \left(\frac{\partial V^-}{\partial w_{ji}}\right)_n. \label{eq:svpn-trans}
\end{align}
Since we have $(V^+)_m = (V^-)_m$ for all other, non-spiking neurons $m\neq n$, it holds that
\begin{align}
	\left(\frac{\partial V^+}{\partial w_{ji}}\right)_m + (\dot V^+)_m \taupost{} = \left(\frac{\partial V^-}{\partial w_{ji}}\right)_m + (\dot V^-)_m \taupost{}.
\end{align}
Because the spiking neuron $n$ causes the synaptic current of all neurons $m\neq n$ to jump by $w_{mn}$, we have
\begin{align}
	\taum (\dot V^+)_m = \taum (\dot V^-)_m + w_{mn}
\end{align}
and therefore get with \cref{eq:postdiff}
\begin{align}
	\left(\frac{\partial V^+}{\partial w_{ji}}\right)_m & = \left(\frac{\partial V^-}{\partial w_{ji}}\right)_m - \taum^{-1}w_{mn} \taupost{}                                    \\
	          & = \left(\frac{\partial V^-}{\partial w_{ji}}\right)_m + \frac{1}{\taum(\dot V^-)_n}w_{mn}\left(\frac{\partial V^-}{\partial w_{ji}}\right)_n.\label{eq:svpm-trans}
\end{align}

\paragraph{Synaptic Current Transition}
The spiking neuron $n$ causes the synaptic current of all neurons $m\neq n$ to jump by the corresponding weight $w_{mn}$.
We therefore have
\begin{align}
	(I^+)_m = (I^-)_m + w_{mn}.
\end{align}
By differentiation, this relation implies the consistency equations for the partial derivatives $\frac{\partial I}{\partial w_{ji}}$ with respect to the considered weight $w_{ji}$,
\begin{align}
	\left(\frac{\partial I^+}{\partial w_{ji}}\right)_m + (\dot I^+)_m\taupost{} = \left(\frac{\partial I^-}{\partial w_{ji}}\right)_m + (\dot I^-)_m\taupost{} + \delta_{in} \delta_{jm},
\end{align}
where $\delta_{ji}$ is the Kronecker delta.
Because
\begin{align}
	\taus(\dot I^+)_m = \taus(\dot I^-)_m - w_{mn},
\end{align}
we get with \cref{eq:postdiff}
\begin{align}
	\left(\frac{\partial I^+}{\partial w_{ji}}\right)_m & = \left(\frac{\partial I^-}{\partial w_{ji}}\right)_m + \taus^{-1}w_{mn}\taupost{}+ \delta_{in} \delta_{jm}                                       \\
	          & = \left(\frac{\partial I^-}{\partial w_{ji}}\right)_m - \frac{1}{\taus (\dot V^-)_n} w_{mn}\left(\frac{\partial V^-}{\partial w_{ji}}\right)_n+ \delta_{in} \delta_{jm}. \label{eq:sipmtrans}
\end{align}
With $(I^+)_n = (I^-)_n$ and $(\dot I^+)_n = (\dot I^-)_n$, we have
\begin{align}
	\left(\frac{\partial I^+}{\partial w_{ji}}\right)_n = \left(\frac{\partial I^-}{\partial w_{ji}}\right)_n. \label{eq:sipntrans}
\end{align}

Using the relations of the partial derivatives from \cref{eq:taupost,eq:svpn-trans,eq:svpm-trans,eq:sipmtrans,eq:sipntrans} in the transition equation \cref{eq:xi-def}, we now derive relations between the adjoint variables.
Collecting terms in the partial derivatives and writing the index of the spiking neuron for the $k$th spike as $n(k)$, we have
\begin{align}
	\xi_k & = \bigg[\sum_{m\neq n(k)} \bigg[\taum (\lambda_V^- - \lambda_V^+)_m\left(\frac{\partial V^-}{\partial w_{ji}}\right)_m + \taus(\lambda_I^- - \lambda_I^+)_m\left(\frac{\partial I^-}{\partial w_{ji}}\right)_m -\taus \delta_{in(k)}\delta_{jm}(\lambda_I^+)_m \bigg]\nonumber                                                                                                              \\
	      & \quad + \left(\frac{\partial V^-}{\partial w_{ji}}\right)_{n(k)}\left[ \taum \left(\lambda_V^- - \frac{(\dot V^+)_{n(k)}}{(\dot V^-)_{n(k)}}\lambda_V^+\right)_{n(k)}+\frac{1}{ (\dot V^-)_{n(k)}}\left(\sum_{m\neq n(k)}w_{n(k)m}(\lambda_I^+-\lambda_V^+)_m -\frac{\partial l_\mathrm{p}}{\partial \tpost{k}} + l_V^+ - l_V^-\right)\right] \nonumber \\
	      & \quad + \taus(\lambda_I^- - \lambda_I^+)\left(\frac{\partial I^-}{\partial w_{ji}}\right)_{n(k)}\bigg]\bigg\rvert_{\tpost{k}}.
\end{align}
This form dictates the jumps of the adjoint variables for the spiking neuron $n$ and all other, silent neurons $m$,
\begin{subequations}\label{eqs:adjoint-jumps-derivation}
	\begin{align}
		(\lambda_V^-)_n & = \frac{(\dot V^+)_n}{(\dot V^-)_n}(\lambda_V^+)_n + \frac{1}{\taum (\dot V^-)_n}\left[\sum_{m\neq n} w_{mn}(\lambda_V^+ - \lambda_I^+)_m +\frac{\partial l_\mathrm{p}}{\partial \tpost{k}}+l_V^--l_V^+\right], \label{eq:firing-lambda-jump} \\
		(\lambda_V^-)_m & = (\lambda_V^+)_m,                                                                                                                                                                                                                      \\
		\lambda_I^-     & = \lambda_I^+.
	\end{align}
\end{subequations}
With these jumps, the gradient reduces to
\begin{align}\label{eq:adjoint-gradient}
	\frac{\mathrm{d}\mathcal{L}}{\mathrm{d}w_{ji}} & =-\taus\sum_{k=1}^{\npost} \delta_{in(k)} (\lambda_I)_j \\
	                                               & =-\taus\sum_{\textrm{spikes from }i}(\lambda_I)_j.
\end{align}
\paragraph{Summary}
The free adjoint dynamics between spikes are given by \cref{eqs:adjoint-dynamics-derivation} while spikes cause jumps given by \cref{eqs:adjoint-jumps-derivation}.
The gradient for a given weight samples the post-synaptic neuron's $\lambda_I$ when spikes are transmitted across the corresponding synapse (\cref{eq:adjoint-gradient}).
Since we can identify, with $(\dot V^+)_n -(\dot V^-)_n=\taum^{-1}\vartheta$,
\begin{align}
	\frac{(\dot V^+)_n}{(\dot V^-)_n} = \frac{(\dot V^+)_n -(\dot V^-)_n}{(\dot V^-)_n} + 1 = \frac{\vartheta}{\taum (\dot V^-)_n} + 1
\end{align}
the derived solution is equivalent to \cref{eq:loss-gradient,fig:adjoint-system}.

\paragraph{Fixed Input Spikes}
If a given neuron $i$ is subjected to a fixed pre-synaptic spike train across a synapse with weight $w_\textrm{input}$, the transition times are fixed and the adjoint variables do not experience jumps.
The gradient simply samples the neuron's $\lambda_I$ at the times of spike arrival,
\begin{align}
	\frac{\mathrm{d}\mathcal{L}}{\mathrm{d}w_\textrm{input}} = -\taus \sum_{\textrm{input spikes}}(\lambda_I)_i.
\end{align}
\paragraph{Coincident Spikes}
The derivation above assumes that only a single neuron of the recurrent network spikes at a given $\tpost{k}$.
In general, coincident spikes may occur.
If neurons $a$ and $b$ spike at the same time and the times of their respective threshold crossing vary independently as function of $w_{ji}$, the derivation above still holds, with both neuron's $\lambda_V$ experiencing a jump as in \cref{eq:firing-lambda-jump}.